\documentclass{article}
\usepackage{spconf,amsmath,graphicx}
\usepackage[T1]{fontenc}
\usepackage[linesnumbered,ruled]{algorithm2e}
\usepackage{stfloats,color}
\usepackage{bbm}
\usepackage{amsfonts}
\usepackage{amssymb}
\usepackage{cite}
\usepackage{array}
\usepackage{subcaption}
\usepackage{times}
\usepackage{epsfig}
\usepackage{latexsym}
\usepackage{epstopdf}
\usepackage{verbatim}
\usepackage{units}
\usepackage{amsthm}
\usepackage{placeins}
\usepackage{afterpage}
\usepackage{dsfont}
\usepackage{soul}
\usepackage{multicol}
\usepackage{multirow}
\usepackage{mathtools}
\usepackage[cmintegrals]{newtxmath}
\usepackage{url}
\newcolumntype{P}[1]{>{\centering\arraybackslash}p{#1}}
\newcolumntype{M}[1]{>{\centering\arraybackslash}m{#1}}

\newtheorem{lemma}{Lemma}

\newenvironment{Proof}[1]{\medskip\par\noindent{\bf Proof:\,}\,#1}{{\mbox{\,$\blacksquare$}\par}}

\title{Age of Gossip with the Push-Pull Protocol}
\name{Arunabh Srivastava, Thomas Jacob~Maranzatto, Sennur Ulukus}
\address{Department of Electrical and Computer Engineering, 
University of Maryland, MD}

\begin{document}
\ninept
\maketitle

\begin{abstract}
We consider a wireless network where a source generates packets and forwards them to a network containing $n$ nodes. The nodes in the network use the asynchronous push, pull or push-pull gossip communication protocols to maintain the most recent updates from the source. We use the version age of information metric to quantify the freshness of information in the network. Prior to this work, only the push gossiping protocol has been studied for age of information analysis. In this paper, we use the stochastic hybrid systems (SHS) framework to obtain recursive equations for the expected version age of sets of nodes in the time limit. We then show that the pull and push-pull protocols can achieve constant version age, while it is already known that the push protocol can only achieve logarithmic version age. We then show that the push-pull protocol performs better than the push and the pull protocol. Finally, we carry out numerical simulations to evaluate these results.
\end{abstract}
\section{Introduction}\label{sec: introduction}
In recent years, wireless technology has progressed at a rapid pace. Large wireless networks for monitoring various processes are becoming a reality under the internet of things (IoT) paradigm. Many tasks carried out by such networks are time-critical, and freshness of information is becoming an important metric for the functioning of such networks. It is well known that throughput and latency are not enough to quantify freshness of information at the nodes in a network~\cite{popovski2022perspective}. Many new metrics have been designed to quantify freshness of information, such as age of information~\cite{kaul2012real, sun2019age, yatesJSACsurvey}, age of incorrect information~\cite{maatouk20AOII}, age of synchronization~\cite{zhong18AoSync}, binary freshness~\cite{cho3BinaryFreshness}, and version age of information~\cite{yates21gossip, Abolhassani21version, melih2020infocom}.

In this paper, we study the version age of information (age of gossip) metric with respect to three protocols we outline shortly: 1) push-only communication, 2) pull-only communication, and 3) push-pull communication. In the networks we consider, there is a source node that generates updates and shares these updates with nodes in a gossiping network. At any time, the source holds a \textit{version-stamped} packet.  When the source updates node $v$ at time $t$, node $v$ inherits the version the source holds at time $t$.  The source can in general send the same packet to many nodes, or never share a packet with any nodes. The \textit{version age} of a node $v$ in the network is defined as the difference between the version-stamps on the packets the source and $v$ hold. Nodes in the network share packets by randomly querying a neighbor in continuous-time and push or pull (or both) data from this neighbor.  In the next section, we will formally define these protocols.

The stochastic hybrid systems (SHS) framework was used for the first time by Yates~\cite{yates21gossip} to find the version age of nodes in a gossiping network following the push protocol. He also showed that the version age of a single node in the fully connected network scales logarithmically with the size of the network.  Subsequently many authors have focused on studying network topologies other than the fully connected network~\cite{buyukates22ClusterGossip, 
srivastava2024networkconnectivityinformationfreshnesstradeoff, maranzatto24}.  Adversarial models were introduced in~\cite{kaswan22timestomp, kaswan22jamming} to study timestomping and jamming attacks. Distributed protocols were studied in~\cite{mitra_allerton22}. We direct the reader to the recent survey on applying the SHS framework to the analysis of gossip networks~\cite{kaswan2023versionagesurvey}. 

To the best of our knowledge, all previous work on studying freshness in gossip networks have focused on the push-only protocol, i.e., nodes only forward packets to their neighbors.  In this paper, we explore two more gossiping protocols that are popular in the literature, the asynchronous pull protocol and the asynchronous push-pull protocol. The asynchronous push-pull protocol was first introduced in~\cite{Boyd06randomgossip} for averaging aggregation on arbitrary graphs. They found that the aggregation time is a function of the mixing time, and applied this result to random geometric and preferential attachment graphs. There have been many influential works in this area since then\cite{pu2020push, dimakis2010gossip}. The push-pull protocol is known to have many benefits over the push and the pull protocols, such as scalability, robustness and fast convergence. 

\begin{figure}
    \centering
    \includegraphics[width=0.8\linewidth]{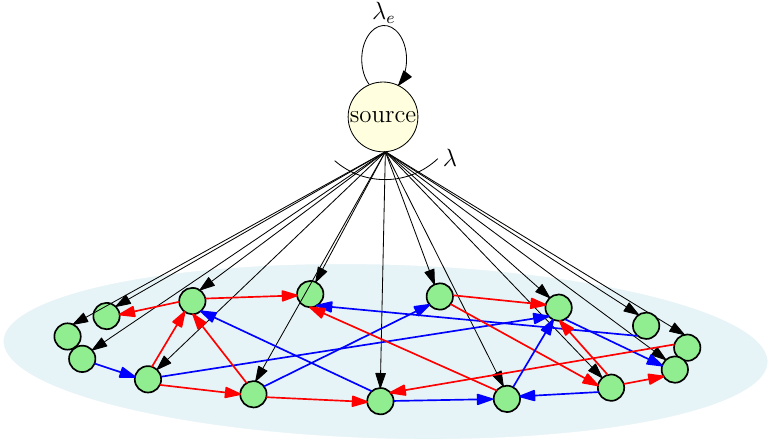}
    \caption{A gossiping network following the push-pull protocol. Each node can push or pull information from neighboring nodes. Arrows denote the flow of information, with red arrows denoting push updates and blue arrows denoting pull updates.}
    \label{fig: push pull main fig}
\end{figure}

In this paper, we extend the use of the SHS framework to find recursive equations for the version age of any subsets of nodes in a network in terms of their one larger supersets. We then highlight the star network with different source updates to emphasize the differences between all three gossiping protocols. We see that even though the push protocol achieves $\Theta(n)$ version age in these networks, the pull and push-pull protocols are capable of achieving $\Theta(1)$ version age. Next, we show that the push-pull protocol performs better than just the push or the pull protocol in an arbitrary gossiping network. Finally, we carry out numerical simulations and compare our theoretical results to empirical observations for the star network. We also analyze the impact of the protocol on version age for the ring network and fully connected network.

\section{System Model and the Version Age Metric}\label{sec: system model}
We consider a source node $n_0$ sending updates to a directed network $G = (\mathcal{N}, E)$ over $n$ nodes. The source updates itself via a Poisson process with rate $\lambda_e$ independent of all other processes in the network. The source also sends updates to each $i \in \mathcal{N}$ as separate, independent Poisson processes with rates $\lambda_{0i}$. Hence, by the thinning of a Poisson Process, the source shares updates with the entire network as a combined rate $\lambda_0 = \sum_{i \in \mathcal{N}}\lambda_{0i}$ Poisson process. For a set of nodes $S \in \mathcal{N}$, we define the total rate at which the source sends information to $S$ as $\lambda_0(S) = \sum_{i \in S} \lambda_{0i}$.

The nodes in $\mathcal{N}$ gossip with each other using the asynchronous push-pull protocol. Any node $i$ pulls the packet stored at node $j$ as a rate $\lambda_{ij}^{\text{pull}}$ Poisson process and pushes its own packet as a rate $\lambda_{ij}^{\text{push}}$ Poisson process. Either rate could be 0 in general, and we say that a node $i$ is a neighboring node of set of nodes $S \subseteq \mathcal{N}$ if $\lambda_{ij}^{\text{pull}} > 0$ or $\lambda_{ij}^{\text{push}}>0$ for some $j \in S$.  We do not require $\lambda_{ij}^{\text{push}} = \lambda_{ji}^{\text{push}}$, and likewise for the pull-edges. Define $\lambda_i^{\text{pull}}(S)$ as the total rate with which nodes in $S$ pulls packets from node $i$. Similarly, define $\lambda_i^{\text{push}}(S)$ as the total rate with which node $i$ pushes packets to nodes in $S$. More formally,
\begin{align}
    \lambda_i^{\text{pull}}(S) = \begin{cases}
        \sum_{j \in S} \lambda_{ji}^{\text{pull}}, & i \notin S\\
        0, & \text{otherwise},
    \end{cases}
\end{align}
and
\begin{align}
    \lambda_i^{\text{push}}(S) = \begin{cases}
        \sum_{j \in S} \lambda_{ij}^{\text{push}}, & i \notin S\\
        0, & \text{otherwise}.
    \end{cases}
\end{align} 
We define $N(S)$ as the set of neighbors of $S$, and $E(S)$ as the set of edges with one endpoint in $S$ and one endpoint in $\Bar{S}$

We use the version age of information metric to quantify the freshness of information at each node. In order to define the version age, we start with the associated counting processes. Let $N_0(t)$ be the counting process associated with the Poisson updates at the source. Similarly, let $N_i(t)$ be the counting process associated with the update versions at $i \in \mathcal{N}$. We note that while the process associated with $N_0(t)$ is Poisson, the process associated with $N_i(t)$ is in general not Poisson; the evolution of this process is discussed in the next paragraph. The version age of $i$ at time $t$ is defined as $X_i(t) = N_0(t) - N_i(t)$. Further, the version age of a set of nodes $S$ in the gossiping network is defined as $X_S(t) = \min_{i \in S}X_i(t)$. We define the limiting average version age of a set $S$ as $v_S = \lim_{t \rightarrow \infty} \mathbb{E}[X_S(t)]$.

The evolution of the version age of a node $i$ in the network by following the push-pull protocol is as follows: If $i$ receives an update from the source, then $i$'s version age falls to $0$, since the source has the latest packet. If the source generates a new packet by updating itself, then the version age of $i$ increases by $1$. If $i$ pulls a packet from node $j$ (because the Poisson process associated to $\lambda_{ij}^{\text{pull}}$ has an arrival), then $i$ keeps whichever packet is fresher between it and node $j$. In a similar way, if node $i$ pushes an update to node $j$, then $j$ updates itself to the fresher packet.  In this context fresher means the packet with larger timestamp, since $n_0$ generates packets with timestamps in $\mathbb{R}^{\geq 0}$.

\section{Recursive Equations and Bounds}\label{sec: recursive equations}
In this section, we use the SHS characterization following \cite{yates21gossip} and \cite{hespanha_SHS} and obtain recursive equations that find the version age of $S$ in terms of sets that contain $S$ and exactly one neighboring node of $S$. Since the rates of all Poisson processes in the network are constant as a function of time and all information exchange processes are memoryless, we have only one discrete state $\mathcal{Q} = \{0\}$, which we now omit from notation. We define the continuous state to be the vector of version ages of nodes in the network, i.e., $\mathbf{X} = [X_1(t), \ldots, X_n(t)]$. Version age of any node in the network is piecewise constant, which implies that the stochastic differential equation is $\Dot{\mathbf{X}}(t) = 0$.

We now define the transition/reset maps $\mathcal{L}$. We can uniquely define each transition map using two variables $(i,j)$ where $i$ denotes the node that is sending information and $j$ denotes the node that is receiving information. Now, there are three types of transitions. The first is when the source updates itself, which we describe using $(0,0)$. The second type of transition is when the source sends updates to a node $i$ in the network. This is denoted using $(0,i)$. The third type of transition is when node $i$ pushes an update to node $j$ or node $j$ pulls an update from node $i$, in which case, the transition is represented as $(i,j)$. This can be summarized as,
\begin{align}
    \mathcal{L} = \{(0,0)\} \cup \{(0,i): i \in \mathcal{N}\} \cup \{(i,j): i,j \in \mathcal{N}\}.
\end{align}
The corresponding transitions can be defined as $\phi_{ij}: \mathbb{R}^n \rightarrow \mathbb{R}^n$ where $\phi_{ij}(\mathbf{X}) = [X_1', X_2', \ldots, X_n']$. Then,
\begin{align}
    X_l' = \begin{cases}
        X_l+1, & i=0,j=0\\
        0, & i=0,j=l \in \mathcal{N}\\
        \min(X_i,X_j) & j=l; i,j \in \mathcal{N}\\
        X_l & \text{otherwise.}
    \end{cases}
\end{align}
The rates of these transitions are then defined as,
\begin{align}\label{def: rates}
    \lambda_{ij} = \begin{cases}
        \lambda_e, & i=0,j=0\\
        \lambda_{0i}, & i=0,j \in \mathcal{N}\\
        \lambda_{ji}^{\text{pull}}+\lambda_{ij}^{\text{push}}, & i,j \in \mathcal{N}.
    \end{cases}
\end{align}
The test functions are chosen to be time variant and dependent on the version age of a set of nodes $S$, $\psi_S(\mathbf{X}) = X_S$. We can now write the extended generator function and evaluate it,
\begin{align}\label{eq: extended generator}
    (L\psi_S)(\mathbf{X}) = \sum_{i,j \in \mathcal{N}}(\psi_S(\phi_{ij}(\mathbf{X}))-\psi_S(\mathbf{X}))\lambda_{ij}.
\end{align}
We evaluate $\psi_S(\phi_{ij}(\mathbf{X}))$, and see that $\psi_S(\phi_{00}(\mathbf{X})) = X_S+1$, $\psi_S(\phi_{0i}(\mathbf{X}))=0, i \in \mathcal{N}$ and $\psi_S(\phi_{ji}(\mathbf{X})) = X_{S \cup \{j\}}, i \in S, j \notin S$. 
We substitute this in \eqref{eq: extended generator}, take expectation on both sides and use Dynkin's formula which results in the left side becoming zero due to version age being piecewise constant. Then, we rearrange it as,
\begin{align}
    \lambda_e =  \lambda_{0}(S)v_S(t) + \sum_{j \in S} \sum_{i \notin S} \left(\lambda_{ji}^{\text{pull}}+\lambda_{ij}^{\text{push}}\right) (v_{S \cup \{i\}}(t)-v_S(t)).
\end{align}
Finally, we take the limit $t \rightarrow \infty$ and rearrange the equation and get the following recursive equations,
\begin{align}\label{eq: recursive equation}
    v_S = \frac{\lambda_e + \sum_{i \in N(S)}\left(\lambda_{i}^{\text{pull}}(S)+\lambda_{i}^{\text{push}}(S)\right)v_{S \cup \{i\}}}{\lambda_0(S) + \sum_{i \in N(S)}\left(\lambda_{i}^{\text{pull}}(S)+\lambda_{i}^{\text{push}}(S)\right)}.
\end{align}
If we have push-only communication, then \eqref{eq: recursive equation} simplifies to the recursive equations found in \cite{yates21gossip}. On the other hand, if we have pull-only communication, then the recursive equations simplify to,
    \begin{align}
        v_S = \frac{\lambda_e + \sum_{i \in N(S)}\lambda_{i}^{\text{pull}}(S)v_{S \cup \{i\}}}{\lambda_0(S) + \sum_{i \in N(S)}\lambda_{i}^{\text{pull}}(S)}.
    \end{align}
We can also write the following lower and upper bounds for $v_S$ following the results in \cite{srivastava2024networkconnectivityinformationfreshnesstradeoff},
\begin{align}
    v_S \leq& \frac{\lambda_e + |N(S)| \cdot \min_{i\in N(S)}{\lambda_i(S)} \cdot \max_{i\in N(S)}{v_{S \cup \{i\}}}}{\lambda_0(S) + |N(S)| \cdot \min_{i\in N(S)}\lambda_i(S)},\\
    v_S \geq& \frac{\lambda_e + |N(S)| \cdot \max_{i\in N(S)}{\lambda_i(S)} \cdot \min_{i\in N(S)}{v_{S \cup \{i\}}}}{\lambda_0(S) + |N(S)| \cdot \max_{i\in N(S)}\lambda_i(S)},
\end{align}
where $\lambda_i(S) = (\lambda_{i}^{\text{pull}}(S)+\lambda_{i}^{\text{push}}(S))$.

\section{A Comparison of the Gossip Protocols}\label{sec: comparison of gossip protocols}
In this section, we evaluate the version age in a star network, and observe how the pull protocol significantly outperforms the push protocol and the push-pull protocol outperforms both the push and the pull protocols in certain cases. Moreover, we see that with the new gossiping protocols, we can beat the best performance of the push-only protocol which occurs in the fully connected network. 

\begin{figure}
    \centering
    \includegraphics[width=\linewidth]{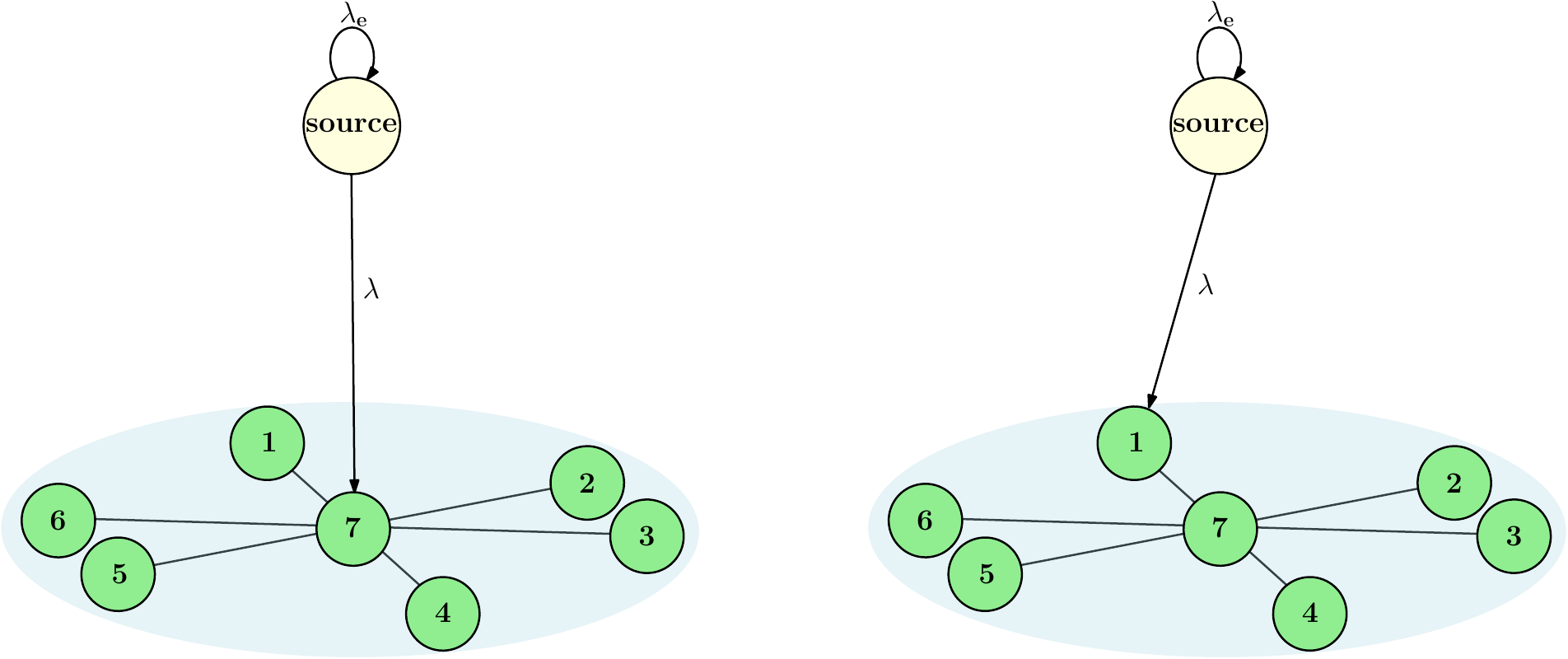}
    \caption{A star network where only one node receives updates directly from the source node. In the left network, the central node receives the updates and in the right network, a non-central node receives the updates. In this example, $n=7$.}
    \label{fig: star network example}
\end{figure}

First, we define the two star networks that we will analyze. These networks are described in Fig.~\ref{fig: star network example}. We assume the central vertex is labelled $n$.  In both networks, all nodes will push/pull packets with total rate $\lambda$ spread evenly over neighbors, so that for $i \not= n$, $\lambda_{in}^{\text{push}} = \lambda$ while $\lambda_{ni}^{\text{push}} = \frac{\lambda}{n-1}$ (and similarly for the pull rates).

In the first star network, shown on the left of Fig.~\ref{fig: star network example}, the source only sends updates to $n$ in the star network, with rate $\lambda$. We can now calculate the version age of a non-central node in the network, say node $1$, since all non-central nodes in the network have the same version age. We do so for the push protocol first,
\begin{align}
    v_1 =& \frac{\lambda_e + \frac{\lambda}{n-1}v_{\{1,n\}}}{\frac{\lambda}{n-1}}\\
    =& (n-1)\frac{\lambda_e}{\lambda} + v_{\{1,n\}}\\
    \geq& \Omega(n)\label{eq: push star network},
\end{align}
due to the non-negativity of version age. Since the upper bound for any push-network is $O(n)$~\cite{maranzatto24}, the average version age of the network with push gossiping scales as $\Theta(n)$. Next, we find $v_1$ for the pull protocol,
\begin{align}
    v_1 =& \frac{\lambda_e + \lambda v_{\{1,n\}}}{\lambda}\\
    =& \frac{\lambda_e}{\lambda}+v_{\{1,n\}}\\
    =& \frac{\lambda_e}{\lambda}+\frac{\lambda_e+\frac{\lambda}{n-1}v_{\{1,2,n\}}}{\lambda+\frac{\lambda}{n-1}}\\
    \leq& \left(2 + O\left(\frac{1}{n}\right)\right)\frac{\lambda_e}{\lambda},
\end{align}
where the last inequality follows from a symmetry argument, i.e., a pull-only network can be thought of as a push-only network with flipped edges. We observe that while both networks have $\Theta(n)$ total communication rate, this rate is more uniform in the push-network, whereas it is heavily concentrated on the central vertex in the pull-network. Thus, the push-pull protocol also performs better than the push protocol, as we will see in Lemma~\ref{lemma: d regular}.

Next, we compare the pull and push-pull protocol using the network illustrated on the right in Fig.~\ref{fig: star network example}. It is easy to see that under the push protocol, for any $i \not= 1$, this network will have worse average version age compared to the previously analyzed network. Thus, we start our calculations with the pull protocol on the central vertex,
\begin{align}
    v_n =& \frac{\lambda_e + \frac{\lambda}{n-1}v_{\{1,n\}}}{\frac{\lambda}{n-1}}\\
    =& (n-1)\frac{\lambda_e}{\lambda}+v_{\{1,n\}}\\
    =&\Omega(n),
\end{align}
in a similar way to \eqref{eq: push star network}. Hence, the version age of node $n$ is $\Theta(n)$. Furthermore, any $j \in \{2,\ldots, n-1\}$ can only access the source's packets through $n$, so every node apart from $1$ has linear age scaling.

Next, we analyze the version age of this network under the push-pull protocol. Once again, any information to nodes $\mathcal{N}' = \{2,\ldots,n-1\}$ passes from node $1$ through node $n$. Hence, all nodes in this set will have worse version age than node $n$ and due to the symmetry of the network, all nodes in this set will have the same limiting average version age. This condition holds for every set containing an equal number of nodes from $\mathcal{N}'$ and either $1$ or $n$ or both. Hence, we define $\{i\}$, $\{1,i\}$, $\{n,i\}$ and $\{1,n,i\}$ to represent sets containing $i$ nodes from $\mathcal{N}'$ and either of $1$ and $n$.

First, we notice that $v_1 = \frac{\lambda_e}{\lambda}$ because $\lambda_0(S) = \lambda \mathbbm{1}\{1 \in S\}$. Hence, it has the minimum version age out of all nodes in any subset of $\mathcal{N}$. This results in any set having node $1$ in it to have $\frac{\lambda_e}{\lambda}$  limiting average version age. Next, we note that for $i < n-2$,
\begin{align}
    v_{\{n,i\}} =& \frac{\lambda_e+(\lambda+\frac{\lambda}{n-1})(v_{\{1,n,i\}}+(n-i-2)v_{\{n,i+1\}})}{(n-i-1)(\lambda+\frac{\lambda}{n-1})} \\
    \leq& \frac{2\lambda_e}{\lambda}\frac{1}{n-i-1}+\frac{n-i-2}{n-i-1}v_{\{n,i+1\}}.
\end{align}
Then, we can write out the recursion as follows,
\begin{align}
    v_{\{1\}} =& \frac{\lambda_e + (\lambda+\frac{\lambda}{n-1})v_{\{n,1\}}}{\lambda+\frac{\lambda}{n-1}}\\
    =& \frac{\lambda_e}{\lambda+\frac{\lambda}{n-1}}+\frac{\lambda_e+(\lambda+\frac{\lambda}{n-1})(v_{\{1,n,1\}}+v_{\{n,2\}})}{2(\lambda+\frac{\lambda}{n-1})}\\
    &\vdots \nonumber\\
    \leq& \frac{\lambda_e}{\lambda} + 2\frac{\lambda_e}{\lambda}\sum_{l=1}^{n-3} \frac{1}{n-2} + o(1)\\
    \leq& 3\frac{\lambda_e}{\lambda}.
\end{align}
Hence, the push-pull protocol is able to achieve constant version age, whereas the pull-only and push-only protocols performs very poorly. These two networks show us that there are networks where the pull protocol performs much better than the push protocol. Further, we observe that the push-pull protocol performs better than both the push and the pull protocols in some networks.

We note that the version age of any gossiping network with the push gossiping protocol cannot be better than $\log{n}$, since the fully connected network has version age $\log{n}$ \cite{yates21gossip}, and it was shown in \cite{maranzatto24} that adding more edges does not decrease version age. On the other hand, both the pull and push-pull gossiping protocols can achieve $O(1)$ version age in certain gossip networks.

\begin{figure}[t]
    \centering
    \includegraphics[width=0.7\linewidth]{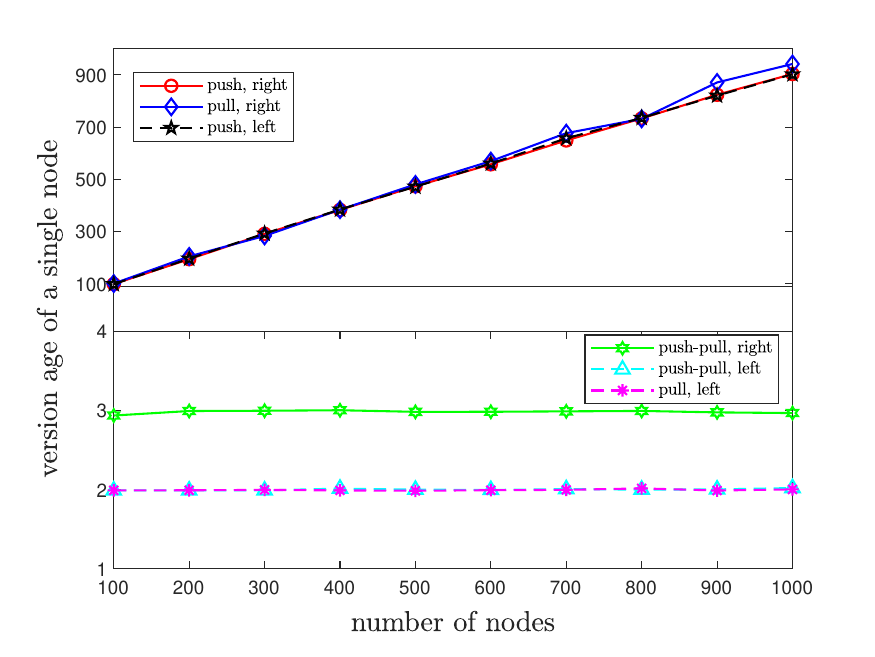}
    \caption{A comparison of the push, pull and push-pull protocols for the star networks, explained in Section~\ref{sec: numerical results}.}
    \label{fig: star network plot}
\end{figure}

\section{Comparison between Push-Pull Protocol and Push-Only/Pull-Only Protocols}\label{sec: d regular}
In this section, we show that with appropriately scaled weights to maintain equal total gossip between two nodes, the push-pull protocol is better than the push-only and pull-only protocols.

\begin{lemma}\label{lemma: d regular}
    For any gossip network $G$,  for any $S \subseteq \mathcal{N}$,
    \[
        v_S^{\text{push-pull}} \leq v_S^{\text{push}} \quad \text{ and } \quad v_S^{\text{push-pull}} \leq v_S^{\text{pull}}
        \]
    where $v_S^{\text{push}}$ is the limiting average version age of $S$ under the push-only  protocol, and likewise for $v_S^{\text{push-pull}}$ and $v_S^{\text{pull}}$.
\end{lemma}
\begin{Proof}
    We verify the first inequality, as the second follows from an identical argument.  Notice that for any edge $(ij)$ in a push-pull network $G$, we can view the forwarding of a packet as a push-only process with a superposition of rates.  In particular, consider a Poisson point process with rate $\lambda_{ij} = \lambda_{ij}^{\text{push}} + \lambda_{ji}^{\text{pull}}$. Then, if $G'$ is a push-only network with edge rates given by $\lambda_{ij}$ for every edge $(ij) \in E(G)$, the version age in $G'$ is identical to the version age in $G$, as the arrival of packets at any vertex $v$ has the same distribution in both graphs.
    
    Define $N'_{ij}(t)$ to be the number of updates on edge $(ij)$ in $G'$ before time $t$, and likewise for $N_{ij}(t)$ for the push-only network over $G$ (that is, the pull-rates in $G$ are set to 0). Now, notice that $N'_i(t)$ stochastically dominates, in the CDF-sense, $N_i(t)$. This is clear since both are homogeneous Poisson processes and the rate for $N'$ is at least that of $N$. Since this is true for every edge, for any vertex $v$ the time it takes a packet to travel from the source to $v$ is stochastically less in $G'$ than in $G$. It follows that for any time $t$ and any vertex $v$, the version age at $v$ in $G'$ stochastically dominates the version age in $G$. Therefore, the limiting average version age for the push-pull network is less than the limiting average version age for the push-only network.
\end{Proof}

Therefore, we have that the push-pull protocol outperforms both the push protocol and the pull protocol. Hence, the version ages of the ring network, fully connected network, grid network, generalized ring networks,
unit hypercube networks and $d$-regular random graphs under the push-pull protocol are upper bounded by the version ages found for the respective networks under the push protocol in \cite{buyukates22ClusterGossip, yates21gossip, srivastava2024networkconnectivityinformationfreshnesstradeoff, maranzatto24}.

\begin{figure}[t]
    \centering
    \includegraphics[width=0.7\linewidth]{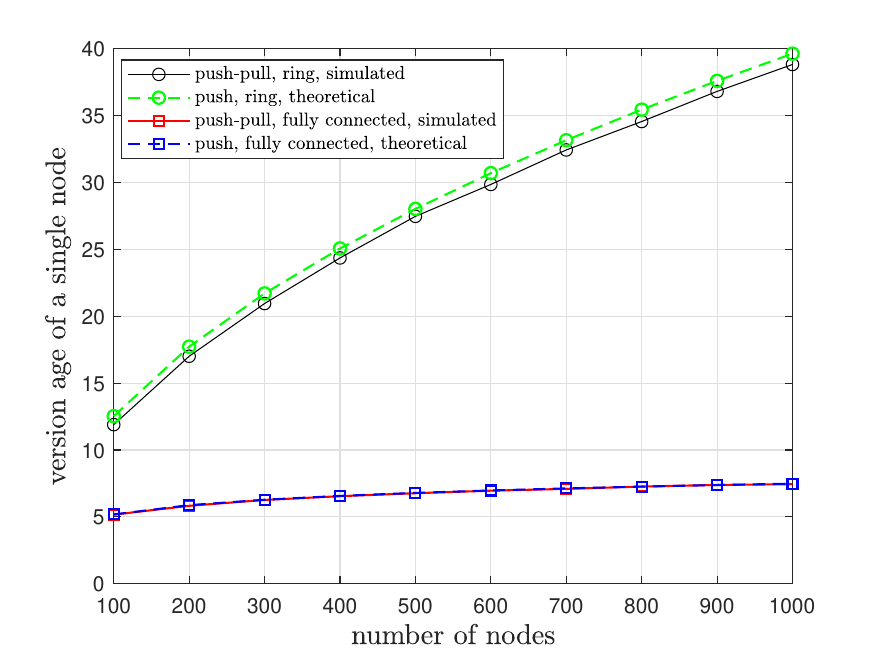}
    \caption{Version age of a single node in the ring and fully connected network with the push-pull protocol when compared to the respective theoretical values under the push protocol found in \cite{buyukates22ClusterGossip} and \cite{yates21gossip}.}
    \label{fig: ring and fc plot}
\end{figure}

\section{Numerical Results} \label{sec: numerical results}
We first verify the results found in Section~\ref{sec: comparison of gossip protocols}. We choose $\lambda_e = \lambda = 1$. We vary the number of users in the star network from $100$ to $1000$ with steps of $100$. We compare how the push, pull and push-pull protocols perform relative to each other with respect to the star graphs from Section~\ref{sec: comparison of gossip protocols}. We see that the bounds we found in Section~\ref{sec: comparison of gossip protocols} are accurate. The push-pull protocol and pull protocol have constant version age for the left star network in Fig.~\ref{fig: star network example}. For the right star network, the push and pull protocols have linear version age while the push-pull protocol has constant version age. 

Next, we simulate the ring network with the push-pull protocol in Fig.~\ref{fig: ring and fc plot}, where $\lambda_e = 1, \lambda^{push} = \lambda^{pull} = 0.5$ for both the push and pull rates for all gossip connections, in order to have a fair comparison with the theoretical bound $\sqrt{\frac{\pi}{2}}\sqrt{n}$ obtained for the ring network in \cite{buyukates22ClusterGossip} with the push protocol. We also simulate the fully connected network with half rate push-pull protocol and compare with the theoretical bound $\log{n}$ found in \cite{yates21gossip}. We see that the theoretical bounds agree with the simulation results and the result in Lemma~\ref{lemma: d regular}.

\section{Conclusion}\label{sec: conclusion}
We considered gossip networks with the pull and push-pull protocol and analyzed them with the version age metric. We found recursive equations to quantify the version age of sets of nodes in the network. We used these recursive equations and the star network to compare the performances of the push, pull and push-pull protocols and showed that $O(1)$ version age is achievable for the pull and push-pull protocol. We then showed that the push-pull protocol outperforms both the push and pull protocols for minimizing the version age. Finally, we verified our results with numerical simulations.

\newpage
\bibliographystyle{IEEEbib}
\bibliography{strings,refs}

\begin{thebibliography}{10}

\bibitem{popovski2022perspective}
P.~Popovski, F.~Chiariotti, K.~Huang, A.~E. Kal{\o}r, M.~Kountouris, N.~Pappas, and B.~Soret,
\newblock ``A perspective on time toward wireless 6{G},''
\newblock {\em Proceedings of the IEEE}, vol. 110, no. 8, pp. 1116--1146, August 2022.

\bibitem{kaul2012real}
S.~K. Kaul, R.~D. Yates, and M.~Gruteser,
\newblock ``Real-time status: How often should one update?,''
\newblock in {\em IEEE Infocom}, March 2012.

\bibitem{sun2019age}
Y.~Sun, I.~Kadota, R.~Talak, and E.~Modiano,
\newblock ``Age of information: A new metric for information freshness,''
\newblock {\em Synthesis Lectures on Communication Networks}, vol. 12, no. 2, pp. 1--224, December 2019.

\bibitem{yatesJSACsurvey}
R.~D. Yates, Y.~Sun, D.~R. Brown, S.~K. Kaul, E.~Modiano, and S.~Ulukus,
\newblock ``Age of information: An introduction and survey,''
\newblock {\em IEEE Jour. on Selected Areas in Communications}, vol. 39, no. 5, pp. 1183--1210, May 2020.

\bibitem{maatouk20AOII}
A.~Maatouk, S.~Kriouile, M.~Assaad, and A.~Ephremides,
\newblock ``The age of incorrect information: A new performance metric for status updates,''
\newblock {\em IEEE/ACM Trans. on Networking}, vol. 28, no. 5, pp. 2215--2228, October 2020.

\bibitem{zhong18AoSync}
J.~Zhong, R.~D. Yates, and E.~Soljanin,
\newblock ``Two freshness metrics for local cache refresh,''
\newblock in {\em IEEE ISIT}, June 2018.

\bibitem{cho3BinaryFreshness}
J.~Cho and H.~Garcia-Molina,
\newblock ``Effective page refresh policies for web crawlers,''
\newblock {\em ACM Trans. on Database Systems}, vol. 28, no. 4, pp. 390--426, December 2003.

\bibitem{yates21gossip}
R.~D. Yates,
\newblock ``The age of gossip in networks,''
\newblock in {\em IEEE ISIT}, July 2021.

\bibitem{Abolhassani21version}
B.~Abolhassani, J.~Tadrous, A.~Eryilmaz, and E.~Yeh,
\newblock ``Fresh caching for dynamic content,''
\newblock in {\em IEEE Infocom}, May 2021.

\bibitem{melih2020infocom}
M.~Bastopcu and S.~Ulukus,
\newblock ``Who should {G}oogle {S}cholar update more often?,''
\newblock in {\em IEEE Infocom}, July 2020.

\bibitem{buyukates22ClusterGossip}
B.~Buyukates, M.~Bastopcu, and S.~Ulukus,
\newblock ``Version age of information in clustered gossip networks,''
\newblock {\em IEEE Jour. on Selected Areas in Information Theory}, vol. 3, no. 1, pp. 85--97, March 2022.

\bibitem{srivastava2024networkconnectivityinformationfreshnesstradeoff}
A.~Srivastava and S.~Ulukus,
\newblock ``Network connectivity--information freshness tradeoff in information dissemination over networks,''
\newblock May 2024,
\newblock Available online at arXiv:2405.19310.

\bibitem{maranzatto24}
T.~J. Maranzatto,
\newblock ``Age of gossip in random and bipartite networks,''
\newblock in {\em IEEE ISIT}, July 2024.

\bibitem{kaswan22timestomp}
P.~Kaswan and S.~Ulukus,
\newblock ``Susceptibility of age of gossip to timestomping,''
\newblock in {\em IEEE ITW}, November 2022.

\bibitem{kaswan22jamming}
P.~Kaswan and S.~Ulukus,
\newblock ``Age of gossip in ring networks in the presence of jamming attacks,''
\newblock in {\em Asilomar Conference}, October 2022.

\bibitem{mitra_allerton22}
P.~Mitra and S.~Ulukus,
\newblock ``{ASUMAN}: Age sense updating multiple access in networks,''
\newblock in {\em Allerton Conference}, September 2022.

\bibitem{kaswan2023versionagesurvey}
P.~Kaswan, P.~Mitra, A.~Srivastava, and S.~Ulukus,
\newblock ``Age of information in gossip networks: A friendly introduction and literature survey,''
\newblock December 2023,
\newblock Available online at arXiv:2312.16163.

\bibitem{Boyd06randomgossip}
S.~Boyd, A.~Ghosh, B.~Prabhakar, and D.~Shah,
\newblock ``Randomized gossip algorithms,''
\newblock {\em IEEE Transactions on Information Theory}, vol. 52, no. 6, pp. 2508--2530, June 2006.

\bibitem{pu2020push}
S.~Pu, W.~Shi, J.~Xu, and A.~Nedi{\'c},
\newblock ``Push--pull gradient methods for distributed optimization in networks,''
\newblock {\em IEEE Transactions on Automatic Control}, vol. 66, no. 1, pp. 1--16, January 2021.

\bibitem{dimakis2010gossip}
A.~G. Dimakis, S.~Kar, J.~M.~F. Moura, M.~G. Rabbat, and A.~Scaglione,
\newblock ``Gossip algorithms for distributed signal processing,''
\newblock {\em Proceedings of the IEEE}, vol. 98, no. 11, pp. 1847--1864, August 2010.

\bibitem{hespanha_SHS}
J.~Hespanha,
\newblock ``Modeling and analysis of stochastic hybrid systems,''
\newblock {\em IEEE Proceedings -- Control Theory and Applications}, vol. 153, no. 5, pp. 520--535, January 2006.

\end{thebibliography}

\end{document}